\documentclass[letterpaper, 10pt, conference]{ieeeconf}
\IEEEoverridecommandlockouts  

\usepackage{amssymb}
\usepackage{amsmath}
\usepackage{amsfonts}                                
\usepackage[latin1] {inputenc}
\usepackage{enumerate}
\usepackage{mathrsfs}
\usepackage{tabulary}
\usepackage{cite}
\usepackage{psfrag}
\usepackage{cite}
\usepackage{graphicx}          
\usepackage{color}      
\usepackage{epsfig} 
\usepackage{epstopdf}
\usepackage{times} 
\def\qedp{\hspace*{\fill}~{\tiny $\blacksquare$}}
\def\qed{\relax\ifmmode\hskip2em \Box\else\unskip\nobreak\hskip1em $\Box$\fi}

\newtheorem{theorem}{Theorem}
\newtheorem{itlemma}{Lemma}
\newtheorem{itdefinition}{Definition}
\newtheorem{itproposition}{Proposition}
\newtheorem{itresult}{Result}
\newtheorem{itremark}{Remark}
\newtheorem{itassumption}{Assumption}
\newtheorem{itcorollary}{Corollary}
\newtheorem{itexample}{Example}

\newenvironment{remark}{\begin{itremark}\rm}{\end{itremark}}
\newenvironment{assumption}{\begin{itassumption}\rm}{\end{itassumption}}
\newenvironment{lemma}{\begin{itlemma}\rm}{\end{itlemma}}

\title{\LARGE \bf Resilient Control under Denial-of-Service: \\ Robust Design} 

\author{Shuai Feng and Pietro Tesi
	\thanks{Shuai Feng and Pietro Tesi are with 
		ENTEG, Faculty of Mathematics and Natural Sciences, University of Groningen, 9747 AG Groningen, The Netherlands
		{\tt\small s.feng@rug.nl, p.tesi@rug.nl}. }
}

\begin{document}
\maketitle           
\begin{abstract}
In this paper, we study 
networked control systems in the presence of Denial-of-Service (DoS) attacks, namely attacks that prevent transmissions over the communication network. 
The control objective is to maximize frequency and duration of the DoS attacks under which closed-loop stability is not destroyed. 
Analog and digital predictor-based controllers with state resetting are proposed, which achieve the considered control objective
for a general class of DoS signals.
An example is given to illustrate 
the proposed solution approach. 
\end{abstract}
	
\section{Introduction}
Owing to advances in computing 
and communication technologies, recent years witnessed 
a growing interest towards cyber-physical systems (CPSs), 
\emph{i.e.}, systems where physical processes are 
monitored/controlled via embedded computers and
networks, possibly with feedback loops that are implemented  
on wireless platforms \cite{lee2008cyber,sha}.
The concept of CPSs is certainly appealing 
for industrial process automation; however, it raises many theoretical
and practical challenges. In particular, the concept of CPSs
has triggered considerable attention towards networked control in the presence of cyber attacks. 
In fact, unlike general-purpose computing systems where attacks limit their impact to the cyber realm, attacks 
to CPSs can affect the physical world: 
if the process under control is open-loop unstable, failures in the plant-controller communication can result in environmental damages. 

The concept of cyber-physical security mostly concerns security against malicious attacks. There are varieties of attacks such as \emph{Denial-of-Service attacks, zero-dynamics attacks, bias injection attacks}, to name a few \cite{Teixeira2015135}. 
The last two are examples of attacks affecting the integrity of data, 
while Denial-of-Service attacks are meant to 
compromise the availability of data.

This paper is concerned with Denial-of-Service (DoS) attacks. 
We consider a sampled-data control system in which the 
measurement channel (sensor-to-controller channel)
is networked; the attacker objective is to induce
closed-loop instability by interrupting  the plant-controller communication. 
In wireless networks, this can be caused by emitting intentional noise,
also known as \emph{jamming}, examples being
constant, random and protocol-aware jamming
\cite{pelechrinis2011denial,tague2009mitigation,debruhl2011digital}. 
It is generally accepted that 
communication failures induced by DoS 
can have a temporal profile quite different 
from the one exhibited by genuine packet losses, 
as assumed in the majority of studies on networked control;
in particular, communication failures induced by DoS 
need not follow a given class of probability distributions \cite{sastry}.
This raises new theoretical challenges from the perspective 
of analysis as well as control design.

In the literature, several contributions 
have been proposed dealing with networked control 
under DoS. In \cite{sastry,basar}, the authors consider the problem of finding optimal control and attack strategies assuming a maximum number of jamming actions over a prescribed (finite) control horizon. A similar 
formulation is considered in \cite{Ugrinovskii}, where the authors 
study zero-sum games between controllers and  strategic jammers.
In \cite{martinez,martinez2},  the authors consider DoS attacks in the form of pulse-width modulated signals. 
The goal is to identify salient features of the DoS signal such as maximum on/off cycle in order to suitably schedule the transmission times. For the case of periodic jamming (of unknown period and duration), 
identification schemes are proposed for de-synchronizing
the transmission times from the DoS signal. 

In \cite{CDP:PT:IFAC14,de2015input}, a framework is introduced where 
no assumption is made regarding the DoS attack underlying strategy. 
A general attack model is considered that only constrains the attacker action in time by posing limitations on the \emph{frequency} of DoS attacks 
and their \emph{duration}. 
The main contribution is an explicit characterization of 
frequency and duration of the DoS attacks 
under which closed-loop stability can be preserved
by means of state-feedback policies. Building on the results in \cite{CDP:PT:IFAC14},
extensions have been considered 
dealing with dynamic controllers \cite{dolk}, 
nonlinear \cite{CDP:PT:CDC14} and distributed \cite{senejohnny} systems.
Recently, a similar formulation has been adopted
in the context of DoS-resilient event-triggered control \cite{Ishii};
see also \cite{dolk}.

From the perspective of securing robustness against DoS, 
static feedback has inherent limitations. 
In fact, using static feedback one 
generates control updates only when new measurements become available.
Intuitively,
this limitation can be overcome by considering 
dynamic controllers. In particular, a natural approach is to
equip the control system with prediction capabilities so as reconstruct the missing measurements from available data during 
the DoS periods. 
Prompted by the above considerations, this paper 
discusses the design of predictor-based controllers 
in the context of DoS-resilient networked control.
Inspired by recent results on finite-time state observers \cite{raff,ferrante}, 
we focus the attention on impulsive-like predictors 
consisting of dynamical observers with 
measurements-triggered state resetting.
Both analog and digital implementations are discussed,
and compared.

While the idea of using predictor-based controllers is
intuitive, the result is perhaps surprising. 
In fact, this paper shows that impulsive-like predictors make it 
possible to 
\emph{maximize} the amount of DoS that one can tolerate 
for the class of DoS signals introduced in \cite{CDP:PT:IFAC14,de2015input}.

\vspace{0.3cm}

The paper is organized as follows. In Section \uppercase\expandafter{\romannumeral 2}, we 
describe the framework of interest, and outline 
the paper contribution. 
Section \uppercase\expandafter{\romannumeral 3} presents the main results. We first design  analog  predictor-based controllers and discuss the conditions under which stability is guaranteed. Second, we design digital predictor-based controllers and characterize sampling rate of the digital device and stability conditions. In Section \uppercase\expandafter{\romannumeral 4}, 
an example is discussed. 
Section \uppercase\expandafter{\romannumeral 5} ends the paper 
with concluding remarks and possible extensions to 
the present research.

\subsection{Notation}
We denote by $\mathbb R$ the set of reals. Given 
$\alpha \in \mathbb R$, we let $\mathbb R_{> \alpha}$
($\mathbb R_{\geq \alpha}$) denote
the set of reals greater than (greater than or equal to) $\alpha$.
We let $\mathbb N_0$ denote the set of nonnegative integers, 
$\mathbb N_0 := \{0,1,\ldots\}$. The prime denotes transpose.
Given a vector $v \in \mathbb R^n$, $\|v\|$ is its Euclidean norm. 
Given a matrix $M$, $\|M\|$ is its spectral norm. 
Given two sets $A$ and $B$,
we denote by $B \backslash A$ the relative complement of $A$ in $B$, 
\emph{i.e.}, the set of all elements belonging to $B$, but not to $A$. 
Given a measurable time function 
$f:  \mathbb R_{\geq 0} \mapsto  \mathbb R^n$ 
and a time interval $[0,t)$
we denote the $\mathcal L_\infty$ norm of $f(\cdot)$ on $[0,t)$ 
by $\|f_t\|_{\infty} := \textrm{sup}_{s \in [0,t)} \|f(s)\|$. 
Given a measurable time function 
$f:  \mathbb R_{\geq 0} \mapsto  \mathbb R^n$ we say 
that $f$ is bounded if its $\mathcal L_\infty$ norm is finite.

\section{The framework}\label{sec.problem}

\subsection{Process dynamics and network}

The process to be controlled 
is given by 
{\setlength\arraycolsep{2pt} 
	\begin{eqnarray} \label{system}
	\left\{ \begin{array}{rl}
	\dot x(t) & = A x(t) + B u(t) + d(t) \\
	y(t) & = x(t) + n(t) \\
	x(0) &= x_0
	\end{array} \right.
	\end{eqnarray}}%
where $t \in \mathbb R_{\geq 0}$; $x \in \mathbb R^{n}$ is the state,
$u \in \mathbb R^{m}$ is the control input and
$y \in \mathbb R^{p}$ is measurement vector; $A$ and $B$ are 
matrices of appropriate size with $(A,B)$ is stabilizable; $d \in \mathbb R^{n}$ and $n \in \mathbb R^{p}$ are unknown (bounded) disturbance and noise
signals, respectively. 

We assume that 
the measurement channel is networked
and subject to Denial-of-Service (DoS) status.
The former implies that measurements are sent 
only at discrete time instants. Let 
$\{t_k\}_{k \in \mathbb N_0} = \{t_0,t_1,\ldots\}$
denote the sequence of transmission attempts. 
Throughout the paper, we assume for simplicity 
that the transmission attempts are carried out periodically 
with period $\Delta$, \emph{i.e.},
\begin{eqnarray} \label{eq:transmission_attempts}
t_{k+1} -t_{k}=\Delta, \quad k \in \mathbb N_0
\end{eqnarray}
with $t_0=0$ by convention. The more general case of
aperiodic transmission policies can be pursued along the
lines of \cite{de2015input}.
We refer to DoS as the phenomenon for which 
some transmission attempts may fail. 
	In this paper, we do not distinguish between transmissions 
	that fail due to channel unavailability 
	(\emph{e.g.}, caused by radio-frequency jammers 
	in protocols employing carrier sensing as medium access policy)
	and transmissions that fail due to DoS-induced packet corruption.
	
	We shall denote by $\{z_m\}_{m \in \mathbb N_0} = \{z_0,z_1,\ldots\}$, 
	$z_0 \geq t_0$,  
    the sequence of time instants at which samples of $y$
    are successfully transmitted. 

\subsection{Control objective} 

The objective is to design $\Delta$ and  
a controller $\mathcal K$, possibly dynamic, in such a way that
the closed-loop stability is maintained
despite the occurrence of DoS periods. 
In this paper,  
by closed-loop stability we mean that 
all the signals in the closed-loop system remain bounded for any initial condition $x_0$ and bounded 
noise and disturbance signals, 
and converge to zero
in the event that noise and disturbance signals 
converge to zero. 

\subsection{Assumptions $-$Time-constrained DoS} \label{sub:DoS}

Clearly, the problem in question does not have a solution 
if the DoS amount is allowed to be arbitrary.
Following \cite{de2015input}, we consider a general DoS model
that constrains the attacker action in time 
by only posing limitations on the frequency of DoS attacks and their duration.
Let 
$\{h_n\}_{n \in \mathbb N_0}$, $h_0 \geq 0$, denote the sequence 
of DoS \emph{off/on} transitions, \emph{i.e.},
the time instants at which DoS exhibits 
a transition from zero (transmissions are possible) to one 
(transmissions are not possible).
Hence,
\begin{eqnarray}  \label{DoS_intervals}
H_n :=\{h_n\} \cup [h_n,h_n+\tau_n[  
\end{eqnarray}
represents the $n$-th DoS time-interval, of a length 
$\tau_n \in \mathbb R_{\geq 0}$,
over which the network is in DoS status. If $\tau_n=0$, then
$H_n$ takes the form of a single pulse at $h_n$.  
Given $\tau,t \in \mathbb R_{\geq0}$ with $t\geq\tau$, 
let $n(\tau,t)$
denote the number of DoS \emph{off/on} transitions
over $[\tau,t[$, and let 
\begin{eqnarray}  \label{DoS_intervals_union}
\Xi(\tau,t) := \bigcup_{n \in \mathbb N_0} H_n  \, \bigcap  \, [\tau,t] 
\end{eqnarray}
denote the 
subset of $[\tau,t]$ where the network is in DoS status. 

We make the following assumptions. 

\begin{assumption}
	(\emph{DoS frequency}). 
	There exist constants 
	$\eta \in \mathbb R_{\geq 0}$ and 
	$\tau_D \in \mathbb R_{> \Delta}$ such that
\begin{eqnarray} \label{ass:DoS_slow_frequency} 
n(\tau,t)  \, \leq \,  \eta + \frac{t-\tau}{\tau_D}
\end{eqnarray}
	for all  $\tau,t \in \mathbb R_{\geq0}$ with $t\geq\tau$.
	\qedp
\end{assumption}

\begin{assumption} 
	(\emph{DoS duration}). 
	There exist constants $\kappa \in \mathbb R_{\geq 0}$ and $T  \in \mathbb R_{>1}$ such that
\begin{eqnarray} \label{ass:DoS_slow_duration}
|\Xi(\tau,t)|  \, \leq \,  \kappa + \frac{t-\tau}{T}
\end{eqnarray}
	for all  $\tau,t \in \mathbb R_{\geq0}$ with $t\geq\tau$. 
	\qedp
\end{assumption}

\begin{remark}
	The rationale behind 
	Assumption 1 is that
	occasionally DoS can occur at a rate faster than $\Delta$
	but the average interval between consecutive DoS triggering is greater than 
	$\Delta$.
	By (\ref{ass:DoS_slow_frequency}), one may in fact have intervals 
	where $h_{n+1} - h_n \leq \Delta$, 
	hence intervals where $n(\tau,t)$ is greater than or equal to the maximum
	number $\lceil (t-\tau)/ \Delta \rceil$ of transmission attempts  
	that may occur within $[\tau,t[$. 
	However, over large time windows, \emph{i.e.}, when the 
	term $(t-\tau)/\tau_D$
	is predominant compared to $\eta$, the number of DoS triggering
	is at most of the order of  $(t-\tau)/\tau_D$.
	Assumption 2 expresses a similar 
	requirement with respect to the DoS duration. In fact, it
	expresses the property that, on the average,
	the time instants over which communication is 
	interrupted do not exceed a certain \emph{fraction} of time,
	as specified by the constant $T \in \mathbb R_{>1}$.
	Similarly to $\eta$,  the constant $\kappa \in \mathbb R_{\geq 0}$ plays the role
	of a regularization term. It is needed because
	during a DoS interval, one has $|\Xi(h_n,h_n+\tau_n)| = \tau_n >  \tau_n /T$
	since $T>1$. Accordingly, $\kappa$ serves to make (6) consistent. 
	Assumptions 1 and 2 are general enough
	to capture many different types of DoS attacks, including
	\emph{trivial, periodic, random} and \emph{protocol-aware jamming} attacks \cite{tague2009mitigation,debruhl2011digital};
	see \cite{de2015input} for a more detailed discussion. \qedp 
\end{remark}
 
\begin{remark}
Unless other conditions are imposed,
both the requirements $\tau_D > \Delta$ and $T >1$ are 
necessary in order for the stabilization problem to be well-posed. 
In fact, if $\tau_D = \Delta$ then the DoS signal 
characterized by the pair $(h_n,\tau_n)=(t_k,0)$ satisfies 
Assumptions 1 and 2 with $\eta=1$, $\kappa=0$
and $T=\infty$ but destroys any communication attempt.
Likewise, in case $T = 1$ then the DoS signal 
characterized by $(h_0,\tau_0)=(0,\infty)$ satisfies 
Assumptions 1 and 2 with $\eta=1$, $\kappa=0$
and $\tau_D=\infty$ but destroys any communication attempt.
\qedp
\end{remark} 
 
\subsection{Previous work and paper contribution}

In \cite{de2015input}, the problem of achieving robustness against DoS 
has been analyzed for the case of
static feedback laws
{\setlength\arraycolsep{2pt} 
	\begin{eqnarray}  \label{eq:static_feedback}
	\arraycolsep=1.4pt\def\arraystretch{1.5}
	u(t) = \left\{ \begin{array}{rl}
	0, & \quad \textrm{} t \in [0,z_0[   \\
     Ky(z_m), & \quad \textrm{} t \in [z_m,z_{m+1}[, \,\, m \in \mathbb N_0
	\end{array} \right.
\end{eqnarray}}%
where $K $ is a state-feedback matrix designed in such a way that all the eigenvalues of $\Phi=A+BK$ have negative real part.
For this scenario, a characterization of 
stabilizing transmission policies was given. 
We summarize below this result.

\begin{theorem} \label{thm:lyap_no_DoS} 
	Consider the process (\ref{system}) under a control action  
	as in (\ref{eq:static_feedback}). 
	Given any positive definite symmetric matrix $Q$,
	let $P$ denote the solution of the Lyapunov equation 
	$\Phi' P + P \, \Phi + Q = 0$. 
	Let the transmission policy in (\ref{eq:transmission_attempts})
	be such that
	\begin{eqnarray} \label{eq:Delta_bar2}
	\Delta  \,  \leq \, 
	\frac{1}{\mu_A} 
	\log \left[ \left( \frac{\sigma}{1+\sigma} \right) \frac{1}{\max\{\|\Phi\|, 
	1\}} \mu_A  +1 \right]
	\end{eqnarray}
	when $\mu_A > 0$, and
	\begin{eqnarray} \label{eq:Delta_bar3}
	\Delta  \,  \leq \, 
	\left( \frac{\sigma}{1+\sigma} \right) \frac{1}{\max\{\|\Phi\|,1\}}
	\end{eqnarray}
	when $\mu_A \leq 0$, where $\mu_A$ is the logarithmic norm of $A$ and 
	$\sigma$ is a positive constant satisfying
	$
	\gamma_1  - \sigma \gamma_2 >0
	$,
	where $\gamma_1$ is equal to the smallest
	eigenvalue of $Q$ and $\gamma_2:=\|2PBK\|$. 
	Then, the closed-loop system is stable
	for any DoS sequence 
	satisfying Assumption 1 and 2 with
	arbitrary $\eta$ and $\kappa$, and with $\tau_D$ and $T$ such that
	\begin{eqnarray} \label{tau_lyap1}
	\frac{1}{T} + \frac{\Delta}{\tau_D} \, < \,  \frac{\omega_1}{\omega_1+ \omega_2}   
	\end{eqnarray} 
	where $\omega_1 := (\gamma_1  - \gamma_2 \sigma)/2\alpha_2$ and
	$\omega_2 := 2 \gamma_2 / \alpha_1$,
	where $\alpha_1$ and  $\alpha_2$ denote the smallest and largest 
	eigenvalue of $P$, respectively.  \qedp
\end{theorem}

Inequality (\ref{tau_lyap1}) provides an explicit characterization 
of the robustness degree against DoS that static feedback policies can achieve.
This characterization relates the DoS parameters $\tau_D$ and $T$
with the transmission period $\Delta$ and the control system 
parameters via $\omega_1$ and $\omega_2$, which depend 
on choice of the state-feedback matrix $K$.

Clearly, increasing the right-hand side of (\ref{tau_lyap1}) increases
the amount of DoS that the control system can tolerate. 
However, with static feedback it is difficult to obtain large 
values for the right-hand side of (\ref{tau_lyap1}).
The underlying reason is that
static feedback has the inherent limitation 
of generating control updates only 
when new measurements become available,
and this possibly reflects in small values for the right-hand side of (\ref{tau_lyap1}).
Intuitively, this limitation can be overcome 
by equipping the controller with prediction capabilities, with the idea
of compensating DoS by reconstructing the missing measurements from available data.
In the next section, it is shown that using
predictor-based controllers one can achieve
closed-loop stability whenever
\begin{eqnarray} \label{eq:main}
\frac{1}{T}+\frac{\Delta}{\tau_D} <1
\end{eqnarray}
holds true.

While the idea of using predictor-based controllers is intuitive, 
the result is perhaps surprising. In fact,
this is 
the best possible bound that one can achieve 
for DoS signals satisfying Assumption 1 and 2. 
Indeed, if we denote by $\mathcal S(\tau_D,T)$ the class of DoS
signals for which (\ref{eq:main}) is not satisfied, then
$\mathcal S(\tau_D,T)$ does always contain DoS signals for which 
stability is destroyed. Examples are DoS signals characterized by 
$(\tau_D,T)=(\Delta,\infty)$ 
and $(\tau_D,T)=(\infty,1)$; \emph{cf.} Remark 2.

\section{Main results}

In Section III-A, we discuss one technical result which 
is fundamental for the developments of the paper. 
The theoretical analysis for analog predictor-based controllers is presented in Section III-B, while in Section III-C we will 
further extend our work to digital implementations. 

\subsection{Key lemma}

The following lemma relates DoS parameters and 
time elapsing between successful transmissions.

\vspace{0.2cm}

\begin{lemma}
Consider a transmission policy as in (\ref{eq:transmission_attempts}),
along with a DoS signal satisfying Assumption 1 and 2. 	
If (\ref{eq:main}) holds true, then the sequence of successful transmissions
satisfies $z_0 \leq Q$ and 
$z_{m+1}-z_{m} \leq Q + \Delta$ for all $m \in \mathbb N_0$, 
where
\begin{eqnarray}
Q=(\kappa + \eta \Delta) \left(1-\frac{1}{T} - \frac{\Delta}{\tau _D} \right)^{-1}
\end{eqnarray}
\end{lemma}

\vspace{0.2cm}

\emph{Proof}.
We first define some auxiliary quantities. 
Let $\bar H_n :=\{h_n\} \cup [h_n, h_n+\tau _n+\Delta[$
represent the $n$-th DoS interval prolonged by one 
sampling. For any interval $[\tau,t]$, let 
$\bar \Xi (\tau,t) := \bigcup_{n \in \mathbb N_0} \bar H_n \bigcap [\tau, t]$
and  
$\bar \Theta (\tau,t):=[\tau,t] \backslash \bar \Xi (\tau,t)$.
The main idea for the proof relies on the following argument.
Given $h_n$, we have
{\setlength\arraycolsep{2pt}
\begin{eqnarray}
|\bar \Theta(h_n,t)| &=& t-h_n - |\bar \Xi(h_n,t)| \nonumber \\
&\geq& t-h_n - |\Xi(h_n,t)| - n(h_n,t)  \Delta  \nonumber \\
&\geq&  ( t-h_n ) \left( 1 - \frac{1}{T} - \frac{\Delta}{\tau_D} \right) -\kappa
- \eta\Delta  
\end{eqnarray}}%
for all $t \geq h_n$
where the first inequality follows from the definition 
of the set $\bar \Xi(\tau,t)$ while the second inequality 
follows from Assumption 1 and 2. 
Notice that $|\bar \Theta (h_n,t)|>0$ implies that  
$[h_n,t]$ contains at least one successful transmission. 
This is because $|\bar \Theta (h_n,t)|>0$ implies that
$[h_n,t]$ contains a DoS-free interval of length greater than $\Delta$.
We claim that 
a successful transmission does always occur 
within $[h_n,h_n+Q]$. To this end, suppose that 
the claim is false and let $t_*$ denote the last 
transmission attempt occurring within $[h_n,h_n+Q]$.
Since $t_*$ is unsuccessful $ |\bar \Theta(h_n,t_*)|=0$.
Moreover, this also implies $ |\bar \Theta(h_n,t_*+\Delta)|=0$.
This is because, if $t_*$ is unsuccessful then it must be contained 
in a DoS interval, say $H_q$, so that 
$[t_*,t_*+\Delta[ \subseteq \bar H_q$.
However, since $t_*+\Delta > h_n+Q$ we also have
{\setlength\arraycolsep{2pt}
\begin{eqnarray} \label{eq:L1_1}
&& |\bar \Theta(h_n,t_*+\Delta)| \nonumber \\
&& \qquad \qquad 
>  Q \left( 1 - \frac{1}{T} - \frac{\Delta}{\tau_D} \right) -\kappa
- \eta\Delta  = 0 \nonumber 
\end{eqnarray}}%
which leads to a contradiction.

Based on these arguments,
the proof can be readily finalized. 
Consider first $z_0 \leq Q$. 
If $t_0$ is successful then the claim holds trivially.
Suppose instead that $t_0$ is unsuccessful, \emph{i.e.},
$h_0  = 0$.
By the above arguments we have one successful transmission
no later than $h_0+Q$ and, hence, no later than
$Q$. Consider next $z_{m+1}-z_{m} \leq Q + \Delta$. 
If $z_m+\Delta$ is successful, then the
claim holds trivially. Suppose instead that $z_m+\Delta$ is unsuccessful.
Since $z_m$ is successful a DoS must occur within $]z_m,z_{m}+\Delta]$.
Hence, we must have $h_n \in ]z_m,z_{m}+\Delta]$ for some $n \in \mathbb N_0$.
By the above arguments we have one successful transmission 
no later than $h_n+Q$ and, hence, no later than $z_{m}+Q+\Delta$. \qedp

\begin{remark}
In the absence of DoS, when  
$T=\tau_D=\infty$ and $\kappa=\eta=0$, $Q$ becomes zero.
In fact, in the absence of DoS, Lemma 1 
simply describes the functioning of a standard periodic transmission policy.
 \qedp
\end{remark}

\subsection{Analog predictor-based controller}

The considered predictor-based controller consists of two parts:  
prediction and state-feedback. As for the prediction part, 
we consider an impulsive predictor, 
whose dynamics are given by 
{\setlength\arraycolsep{2pt} 
	\begin{eqnarray}  \label{eq:ct_predictor}
	\arraycolsep=1.4pt\def\arraystretch{1.1}
	\left\{ \begin{array}{rl}
	\dot { \hat{x}}(t) &= A \hat{x}(t) + B u(t), \quad t \neq z_m \\
	\hat x (t) &= y(t), \quad t = z_m 
	\end{array} \right.
\end{eqnarray}}%
with initial condition
{\setlength\arraycolsep{2pt} 
	\begin{eqnarray}  
	\arraycolsep=1.4pt\def\arraystretch{1.1}
	\hat x (0) = \left\{ \begin{array}{rl}
	y(0), \quad \textrm{if } z_0 = 0  \\
     0, \quad \textrm{otherwise} 
	\end{array} \right.
\end{eqnarray}}%
where $t \in \mathbb R_{\geq 0}$ and $m \in \mathbb N_{0}$.
By construction the solution $\hat x$ is
continuous from the right everywhere.

The state-feedback matrix is an 
arbitrary matrix $K$ such that all the eigenvalues of $\Phi=A+BK$ have 
negative real part. Then,
the control input applied to the process (and the predictor) is given by
\begin{eqnarray} \label{eq:state_feedback}
	u(t)=K\hat x(t)
\end{eqnarray}
where $t \in \mathbb R_{\geq 0}$.

The predictor differs from a classical asymptotic observer
due to the measurements-triggered jumps in the state.
The reason for considering an impulsive-like predictor  
rather than an asymptotic one is the following. Let
\begin{eqnarray} \label{eq:error_ct}
e(t):= \hat x(t) -x(t)
\end{eqnarray} 
where $t \in \mathbb R_{\geq 0}$. The process dynamics can be therefore 
expressed as
\begin{eqnarray} \label{eq:system_equiv_form}
\dot x(t)= \Phi x(t)+BKe(t)+d(t)
\end{eqnarray}
where $t \in \mathbb R_{\geq 0}$.
Consider any symmetric positive definite matrix 
$Q$, and let $P$ be the solution of 
the Lyapunov equation
$\Phi' P+P\Phi +Q=0$. Let $V(x)=x' P x$.
Its derivative along the solutions to (\ref{eq:system_equiv_form}), 
satisfies
{\setlength\arraycolsep{2pt} 
\begin{eqnarray} \label{eq:Lyap}
\dot V(x(t)) \leq &-& \gamma _1 \|x(t)\|^2 + \gamma _2 \|x(t)\|\|e(t)\|  \nonumber \\
&+& \gamma _3 \|x(t)\|\|d(t)\|
\end{eqnarray}}%
for all $t \in \mathbb R_{\geq 0}$, 
where $\gamma _1$ is the smallest eigenvalue of $Q$, $\gamma _2 :=\|2PBK\|$ and $\gamma _3 :=\|2P\|$. 
From the last expression one sees that 
stability depends on the magnitude of $e$. In this respect,
the dynamics of $e$ obeys
{\setlength\arraycolsep{2pt} 
	\begin{eqnarray}  \label{eq:error_dynamics_ct}
		\arraycolsep=1.4pt\def\arraystretch{1.5}
	\begin{array}{rl}
	\dot e(t) &= Ae(t)-d(t), \quad t \ne z_m \\
	e(t) &= n(t), \quad t=z_m \\
	\end{array} 
\end{eqnarray}}%
where $t \in \mathbb R_{\geq 0}$ and $m \in \mathbb N_{0}$.
One sees from the second equation of 
(\ref{eq:error_dynamics_ct}) that
resetting the predictor state makes it possible to
reset $e$ to a bounded value 
whenever a new measurement becomes
available. In turns, Lemma 1 ensures 
that a resetting does always occur in a finite time. 
These two properties guarantee boundedness of $e$
for all $t \geq z_0$.

In particular, we have the following result.

\vspace{0.2cm}

\begin{lemma}
Consider the process (\ref{system}) 
with predictor-based controller 
(\ref{eq:ct_predictor})-(\ref{eq:state_feedback}) 
under a transmission policy as in (\ref{eq:transmission_attempts}).   
Consider any DoS sequence satisfying Assumption 1 and 2 with arbitrary $\eta$ and $\kappa$, and with $\tau _D$ and $T$ satisfying (\ref{eq:main}). 
Then, there exists a positive constant 
$\rho$ such that
{\setlength\arraycolsep{2pt} 
\begin{eqnarray}
\|e(t)\| \, \leq \, \rho \left\| w_t \right\|_\infty
\end{eqnarray}}%
 for all $t \in \mathbb R_{\geq z_0}$, where 
 $w = \left[d' \,\,n' \right]'$.
\end{lemma}

\vspace{0.2cm}

\emph{Proof}. 
Consider any interval $[z_m,z_{m+1}[$, $m \in \mathbb N_0$.
By (\ref{eq:error_dynamics_ct}), we have
\begin{eqnarray}
e(t)=e^{A(t-z_m)}n(z_m)-\int_{z_m}^t e^{A(t-\tau)}d(\tau)d\tau
\end{eqnarray}
for all $t \in [z_m,z_{m+1}[$. 

Let now $\mu_A$ denote the logarithmic norm of $A$.
If $\mu_A \leq 0$, we obtain 
{\setlength\arraycolsep{2pt}
\begin{eqnarray}
\|e(t)\|  &\le&  \|n(z_m)\| + \|d_t\|_\infty (t-z_m) \nonumber\\
		&\le&    \|n_t\|_\infty + \|d_t\|_\infty (Q+\Delta) 
\end{eqnarray}}%
for all $t \in [z_m,z_{m+1}[$, where the second inequality 
follows from Lemma 1. If instead $\mu_A > 0$, we have 
{\setlength\arraycolsep{1pt}
\begin{eqnarray}
\|e(t)\| 
&\leq&   e^{\mu_A(t-z_m)} \|n(z_m)\| + 
\frac{1}{\mu_A} \left( e^{\mu_A (t-z_m)}-1 \right) \|d_t\|_\infty \nonumber \\
&\leq&   e^{\mu_A(Q+\Delta)} \|n_t\|_\infty + 
\frac{1}{\mu_A} \left(e^{\mu_A (Q+\Delta)}-1 \right) \|d_t\|_\infty \nonumber \\
\end{eqnarray}}%
where the second inequality 
follows again from Lemma 1.
Hence, we conclude that the claim
holds with
\begin{eqnarray}
\rho := 1+Q + \Delta
\end{eqnarray}
if $\mu_A \leq 0$, and with 
\begin{eqnarray}
\rho := \left(1+\frac{1}{\mu _A} \right) e^{\mu_A (Q+\Delta)} 
\end{eqnarray}
if $\mu_A > 0$. \qedp

\vspace{0.2cm}

Exploiting Lemma 2,
we obtain the following stability result 
for analog controller implementations. 

\vspace{0.2cm}

\begin{theorem}
Consider the process (\ref{system}) 
with predictor-based controller 
(\ref{eq:ct_predictor})-(\ref{eq:state_feedback}) 
under a transmission policy as in (\ref{eq:transmission_attempts}).   
Then, the closed-loop system is stable for any DoS
sequence satisfying Assumption 1 and 2 with arbitrary $\eta$ and $\kappa$, and with $\tau _D$ and $T$ satisfying (\ref{eq:main}). 
\end{theorem}

\vspace{0.2cm}

\emph{Proof}. 
Consider the closed-loop dynamics for all 
$t \geq z_0$. Notice that $z_0$ exists finite 
by virtue of Lemma 1. In view of (\ref{eq:Lyap}) 
and Lemma 2, we have
{\setlength\arraycolsep{2pt}
\begin{eqnarray}
\dot V(x(t)) &\le&  - \gamma _1 \|x(t)\|^2 
+ \gamma _4 \|x(t)\| \|w_t\|_\infty   
\end{eqnarray}}%
for all $t \in \mathbb{R}_{\geq z_0}$,
where $\gamma _4 := \gamma _2 \rho +\gamma _3$.

Observe that for any positive real $\beta$, the Young's inequality
yields 
\begin{eqnarray} \label{eq:binomial}
2 \|x(t)\|  \|w_t\|_\infty \, \leq  \, \frac{1}{\beta} \|x(t)\|^2 + \beta \|w_t\|_\infty 
\end{eqnarray}
Using this inequality with $\beta=\gamma_4/\gamma_1$, 
straightforward calculations yield
{\setlength\arraycolsep{2pt}
\begin{eqnarray}
\dot V(x(t)) &\leq& -\omega _1 V(x(t)) + \gamma_5 \|w_t\|^2_\infty   
\end{eqnarray}}%
for all $t \in \mathbb{R}_{\geq z_0}$, where 
$\omega _1:= \gamma _1/(2 \alpha_2)$ and 
$\gamma_5:= \gamma_4 ^2/(2 \gamma _1)$, 
where $\alpha_2$ denotes the largest eigenvalue of $P$.
Accordingly, we obtain
{\setlength\arraycolsep{0pt} 
\begin{eqnarray} 
V(x(t)) \, &\le& \, e^{-\omega_1 (t-z_0)} V(x(z_0))) 
+ \frac{\gamma_5}{\omega_1}  \| w_t\|^2_\infty
\end{eqnarray}}%
for all $t \in \mathbb{R}_{\geq z_0}$.
This shows that $x$ remains bounded because $z_0$ exists finite
in view of Lemma 1. In turns, this implies that 
also $\hat x$ remains bounded. 
Moreover, in the event that disturbance and noise signals
converge to zero, (\ref{eq:error_dynamics_ct}) implies that 
$e$ converges to zero. In turns, (\ref{eq:Lyap}) implies that 
both $x$ and  $\hat x$ also converge to zero. \qedp

\begin{remark}
The considered controller yields quite strong stability properties, 
namely global exponential stability with linear bounds on the map
from the disturbance and noise signals to the process state.
It is also interesting to observe that,
as long as the triplet $(\tau_D,T,\Delta)$ 
satisfies (\ref{eq:main}), $\Delta$ can be chosen 
arbitrarily (though large values of $\Delta$ may affect 
the performance via $\gamma_5$, which depends on $\rho$). 
In particular, in the absence of DoS
when $T = \tau_D = \infty$ and $\kappa = \eta = 0$, then (\ref{eq:main})
is satisfied for any bounded value of $\Delta$. This is due to the controller state 
resetting mechanism.
 \qedp
\end{remark}

\subsection{Digital predictor-based controller}

In this section, we extend the control algorithm to a digital implementation. 
The substantial difference between analog and digital implementations
is that in the latter the control action can be updated only at a finite rate.
Because of this, Lemma 2 does not hold any longer. As we will see, in order to recover 
a boundedness inequality similar to the one in Lemma 2, 
constraints have to be enforced on the sampling rate 
of the digital controller. 

Consider a digital controller with sampling rate 
\begin{eqnarray}
\delta=\frac{\Delta}{b} 
\end{eqnarray}
where $b$  is any positive integer. 
Choosing the controller sampling rate 
as a submultiple of $\Delta$ makes it possible 
to implement the controller as a sampled-data 
version of (\ref{eq:ct_predictor}), which is
synchronized with the network transmission rate. 
Let $A_\delta=e^{A\delta}$ and 
$B_\delta=\int_{0}^{\delta} e^{A \tau} B d\tau$.
The digital predictor is given by
{\setlength\arraycolsep{2pt} 
	\begin{eqnarray}  \label{eq:dt_predictor}
	\arraycolsep=1.4pt\def\arraystretch{1.8}
	\left\{ \begin{array}{l}
	\hat{x}((q+1)\delta) = A_\delta \alpha(q\delta)+B_\delta u(q\delta) \\ 
	\alpha(q\delta) = \left\{ \begin{array}{rl}
	y(q\delta), & \quad \textrm{if } q\delta=z_m \\
	\hat x(q\delta), & \quad \textrm{otherwise} 
	\end{array} \right. \\
	\hat x (0) = 0 
	\end{array} \right.
\end{eqnarray}}%
where $q \in \mathbb N_0$.

The control action is given by
\begin{eqnarray} \label{eq:state_feedback_dt}
 u(q\delta) =K\alpha(q\delta). 
\end{eqnarray}
where $q \in \mathbb N_0$.

Similar to the analog implementation, 
also the digital implementation is equipped with a state resetting 
mechanism. Due to the discrete nature of the 
update equations, the resetting mechanism is implemented 
using an auxiliary variable $\alpha$.

The stability analysis follows the same steps as in the previous case.
Let 
\begin{eqnarray} \label{eq:error_dt}
\phi(t):= \alpha(q\delta ) -x(t)
\end{eqnarray} 
where $t \in I_q :=[q\delta, (q+1)\delta[$, $q \in \mathbb N_0$.
Hence, the process dynamics satisfies
\begin{eqnarray} \label{eq:system_equiv_form_dt}
\dot x(t)= \Phi x(t)+BK\phi(t)+d(t)
\end{eqnarray}
for all $t \in I_q$. 

Given any symmetric positive definite matrix 
$Q$, let $P$ be the solution of 
the Lyapunov equation
$\Phi' P+P\Phi +Q=0$. Let $V(x)=x' P x$.
Its derivative along the solutions to (\ref{eq:system_equiv_form_dt}), 
satisfies
{\setlength\arraycolsep{2pt} 
\begin{eqnarray} \label{eq:Lyap_dt}
\dot V(x(t)) \leq &-& \gamma _1 \|x(t)\|^2 + \gamma _2 \|x(t)\|\|\phi(t)\|  \nonumber \\
&+& \gamma _3 \|x(t)\|\|d(t)\|
\end{eqnarray}}%
for all $t \in I_q$,
where $\gamma _1$ is the smallest eigenvalue of $Q$, $\gamma _2 :=\|2PBK\|$ and $\gamma _3 :=\|2P\|$. 
As in the previous case,
stability depends on the magnitude of $\phi$. In this respect,
the dynamics of $\phi$ satisfies
{\setlength\arraycolsep{2pt} 
	\begin{eqnarray}  \label{eq:error_dynamics_dt}
		\arraycolsep=1.4pt\def\arraystretch{1.5}
	\begin{array}{rl}
	\dot \phi(t) &= - \dot x(t) \\
	&= A \phi(t) - \Phi \alpha(q\delta) - d(t), \quad t \ne z_m \\
	\phi(t) &= n(t), \quad t=z_m \\
	\end{array} 
\end{eqnarray}}%
for all $t \in I_q$.

The differential equation in (\ref{eq:error_dynamics_dt}) 
differs from its analog counterpart in 
(\ref{eq:error_dynamics_ct}) due to the extra term 
$\Phi \alpha(q\delta)$.
Because of this, Lemma 2 breaks down. In order to recover 
a property similar to the one established in Lemma 2, 
constraints have to be enforced on the sampling rate 
of the digital controller. This is consistent with intuition, 
and simply indicates that the rate of control updates 
has to be sufficiently fast. In this respect,
letting $\delta=\Delta/b$ allows to differentiate 
between controller sampling rate and transmission rate,
maintaining $\Delta$ possibly large. 

\vspace{0.2cm}

\begin{lemma}
Consider the process (\ref{system}) 
with predictor-based controller  
(\ref{eq:dt_predictor})-(\ref{eq:state_feedback_dt}) 
under a transmission policy as in (\ref{eq:transmission_attempts}).   
Consider any DoS sequence satisfying Assumption 1 and 2 with arbitrary $\eta$ and $\kappa$, and with $\tau _D$ and $T$ satisfying (\ref{eq:main}). 
Let the controller sampling rate 
	be such that
	\begin{eqnarray} \label{eq:Delta_bar2}
	\delta  \,  \leq \, 
	\frac{1}{\mu_A} 
	\log \left[ \left( \frac{\sigma}{1+\sigma} \right) \frac{1}{\max\{\|\Phi\|, 
	1\}} \mu_A  +1 \right]
	\end{eqnarray}
	when $\mu_A > 0$, and
	\begin{eqnarray} \label{eq:Delta_bar3}
	\delta  \,  \leq \, 
	\left( \frac{\sigma}{1+\sigma} \right) \frac{1}{\max\{\|\Phi\|,1\}}
	\end{eqnarray}
	when $\mu_A \leq 0$, where $\mu_A$ is the logarithmic norm of $A$ and 
	$\sigma$ is a positive constant satisfying
	$
	\gamma_1  - \sigma \gamma_2 >0
	$,
	where $\gamma_1$ is equal to the smallest
	eigenvalue of $Q$ and $\gamma_2:=\|2PBK\|$. 
	Then, there exists a positive constant 
$\tilde \rho$ such that
{\setlength\arraycolsep{2pt} 
\begin{eqnarray} \label{eq:tilde_rho}
\|\phi(t)\| \, \leq \, \sigma \|x(t)\| + \tilde \rho \left\| w_t \right\|_\infty
\end{eqnarray}}%
 for all $t \in \mathbb R_{\geq z_0}$.
\end{lemma}

\vspace{0.2cm}

\emph{Proof}.
Consider any interval $[z_m,z_{m+1}[$, 
$m \in \mathbb N_0$, and any controller sampling instant 
$q\delta \in [z_m,z_{m+1}[$.
The proof is divided into two steps. 
In the first step, we provide an upper bound 
on the error dynamics $\phi$ at the controller sampling time $q\delta$.
Second, we provide an upper bound on the error dynamics $\phi$
between controller inter-samplings. In turns, this provides 
an upper bound on $\phi$ over the whole interval $[z_m,z_{m+1}[$, and,
hence, over $\mathbb R_{\geq z_0}$.

For the sake of convenience, we will
relate a controller update instant $q \delta$ with
a successful transmission instant $z_m$ via the 
expression
{\setlength\arraycolsep{2pt} 
\begin{eqnarray}
q\delta = z_m + p\delta 
\end{eqnarray}}%
where $p \in \mathbb N_0$. This is always 
possible since $\delta=\Delta/b$.

We start by deriving an upper bound on $\phi(q\delta)$.
It is simple to verify that
the dynamics of the variable $\alpha$ 
in the controller equations satisfies 
{\setlength\arraycolsep{2pt} 
\begin{eqnarray}
\alpha(q\delta)  &=&  A_\delta^p  \, \alpha(z_m)  \nonumber \\
&& + \sum_{k=0}^{p-1}  A_\delta^{p-k-1}  B_\delta \, u(z_m + k \delta) 
\end{eqnarray}}%
In fact, between two successful transmissions, $\alpha$
coincides with $\hat x$, which evolves like 
a classical linear time-invariant discrete-time system. 

On the other hand,
{\setlength\arraycolsep{2pt} 
\begin{eqnarray}
x(t) \, &=& \, e^{A (t-z_m) } x(z_m) + \int_{z_m}^t   e^{A (t-\tau) } B u(\tau) d \tau
 \nonumber \\ 
&& +  \, \int_{z_m}^t   e^{A (t-\tau) } d(\tau) d \tau
\end{eqnarray}}%
for all $t \in [z_m, z_{m+1}[$. 

Combining the two expressions, we get
{\setlength\arraycolsep{1pt} 
\begin{eqnarray} 
\phi(q\delta ) &=& \alpha(q\delta ) - x(q\delta ) \nonumber \\
&=& e^{A (q \delta -z_m)}n(z_m)-\int_{z_m}^{q\delta }
e^{A(q\delta -\tau)}d(\tau)d\tau  \nonumber \\
\end{eqnarray}}%
where we exploited 
the relation $A_\delta^p=e^{A p \delta }=e^{A (q \delta -z_m)}$, and
the fact that
{\setlength\arraycolsep{2pt} 
\arraycolsep=1.4pt\def\arraystretch{2}
\begin{eqnarray}
&& \int_{z_m}^{q\delta}   e^{A (q\delta-\tau) } 
B u(\tau) d \tau  \nonumber \\
&& \quad  = \sum_{k=0}^{p-1} \left[ \int_{z_m+k\delta}^{z_m+(k+1) \delta} 
e^{A(q\delta-\tau)} B d\tau  
\right] u(z_m+k\delta) \nonumber \\
&& \quad  = \sum_{k=0}^{p-1}  e^{A \delta (p - k  -1)}
\left[ \int_{0}^{\delta} 
e^{A s} B d s  
\right] u(z_m+k\delta) 
\nonumber \\
&& \quad  = \sum_{k=0}^{p-1}  A_\delta^{p - k  -1}
B_\delta u(z_m+k\delta) 
\end{eqnarray}}%
where the second equality is obtained using the change of variable
$s = z_m + (k+1)\delta -\tau$.

We can now obtain an upper bound on $\phi(q\delta )$.
Specifically, since by hypothesis $q \delta \in [z_{m},z_{m+1}[$, 
we have 
{\setlength\arraycolsep{2pt} 
\begin{eqnarray} 
\|\phi(q\delta)\| &\leq& \rho \|w_{q\delta}\|_\infty
\end{eqnarray}}%
where $\rho$ is defined as in Lemma 2.

We can now provide an upper bound on $\phi$
between controller inter-samplings. 
Let 
{\setlength\arraycolsep{2pt} 
\begin{eqnarray} 
f(t-q\delta):=\int_{q\delta}^{t} e^{A(t-\tau)}  d\tau
\end{eqnarray}}%
Integrating (\ref{eq:error_dynamics_dt})
over the interval $I_q$, we obtain
{\setlength\arraycolsep{3pt} 
\begin{eqnarray} 
\arraycolsep=2.4pt\def\arraystretch{3}
\|\phi(t)\| & \leq & \|e^{A(t-q\delta)} \| \|\phi(q\delta)\|  \nonumber \\
&& + \,  f(t-q\delta) \|d_t\|_\infty
+ f(t-q\delta) \|\Phi\| \|\alpha(q\delta)\| 
 \nonumber \\  
 &\leq&  \hat \rho \, \rho  \|w_t\|_\infty  \nonumber \\ 
&& + \,  f(t-q\delta) \|d_t\|_\infty
+ f(t-q\delta) \|\Phi\| \|\alpha(q\delta)\| 
 \nonumber \\  
 &\leq&   \hat \rho \, \rho    \|w_t\|_\infty  \nonumber \\ 
&& + \,  f(t-q\delta) \|d_t\|_\infty  \nonumber \\  
&& + \, f(t-q\delta) \|\Phi\| ( \|\phi(t)\| + \|x(t)\| ) 
\end{eqnarray}}%
for all $t \in I_q$, where $\hat \rho := \max \{ e^{\mu_A \delta},1 \}$.

Let $\kappa_1 := \max \{\|\Phi\|,1\}$.
Observe that $f(0)=0$ and that $f(t-q \delta)$ is monotonically increasing with $t$. 
Accordingly, any positive real $\delta$ such that
\begin{eqnarray} \label{eq:Delta_bar}
f(\delta) \,  \leq \, \frac{1}{\kappa_1} \frac{\sigma}{(1+\sigma)} \, ,
\end{eqnarray}
ensures (\ref{eq:tilde_rho}) 
with
{\setlength\arraycolsep{2pt} 
\begin{eqnarray} 
\tilde \rho : = \sigma + \hat \rho \rho (1+\sigma) 
\end{eqnarray}}%

We finally derive an explicit expression for $\delta$. If  
$\mu_A>0$, we have 
\begin{eqnarray} \label{}
f(\delta) = \frac{1}{\mu_A} (e^{\mu_A \delta} -1 ) 
\end{eqnarray}
and (\ref{eq:Delta_bar2}) yields the desired result.
If instead $\mu_A\leq0$,
then $f(\delta)\leq \delta$, and
(\ref{eq:Delta_bar3}) yields the desired result. 

This concludes the proof. \qedp

\vspace{0.2cm}

Based on Lemma 3 the following result can be stated,
which provides a natural counterpart of Theorem 2.

\vspace{0.2cm}

\begin{theorem}
Consider the process (\ref{system}) 
with predictor-based controller 
(\ref{eq:dt_predictor})-(\ref{eq:state_feedback_dt}) 
under a transmission policy as in (\ref{eq:transmission_attempts}).   
Let the controller sampling rate be chosen as in Lemma 3.
Then, the closed-loop system is stable for any DoS
sequence satisfying Assumption 1 and 2 with arbitrary $\eta$ and $\kappa$, and with $\tau _D$ and $T$ satisfying (\ref{eq:main}). 
\end{theorem}

\vspace{0.2cm}

\emph{Proof.} 
Consider the closed-loop dynamics for all $t \geq z_0$.
Substituting (\ref{eq:tilde_rho}) into (\ref{eq:Lyap_dt})
yields
{\setlength\arraycolsep{2pt} 
\begin{eqnarray}
\dot V(x(t)) &\le& - (\gamma _1 - \sigma \gamma _2) \|x(t)\|^2 \nonumber \\
&& + (\gamma _2 \tilde \rho  + \gamma _3 ) \|x(t)\| \|w_t\|_\infty  
\end{eqnarray}}%
for all $t \in \mathbb R_{z_0}$,
where $\gamma _1 - \sigma \gamma _2$ is strictly positive 
by construction.
The conclusion is that the proof Theorem 2 carries over
to Theorem 3 with $\gamma_1$ and $\gamma_4$ 
replaced by $\gamma _1 - \sigma \gamma _2$ and 
$\gamma _2 \tilde \rho  + \gamma _3$, respectively.
\qedp 

\vspace{0.2cm}

Compared with the analog implementation, one sees that the digital 
implementation does only require a proper choice 
of the controller sampling rate. On the other hand, 
it achieves the same robustness 
properties of the analog implementation.
By Lemma 3, admissible values for the controller sampling rate
can be explicitly computed from the parameters of 
the control system. 

\begin{figure}[tb]
\begin{center}
\psfrag{x1}{{\tiny $x_1$}}
\psfrag{x2}{{\tiny $x_2$}}
\psfrag{DoS}{{\tiny DoS}}
\psfrag{ud}{{\scriptsize $u(t_k)$}}
\includegraphics[width=0.45 \textwidth]{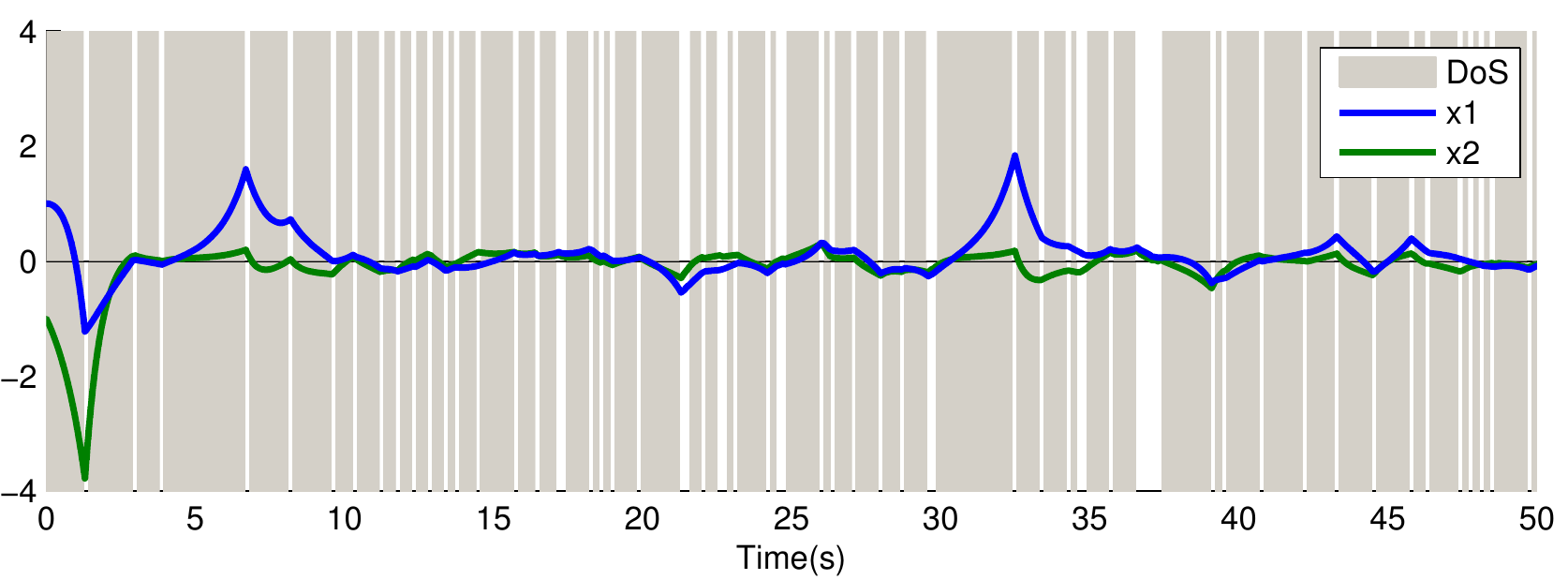} \\
\includegraphics[width=0.45 \textwidth]{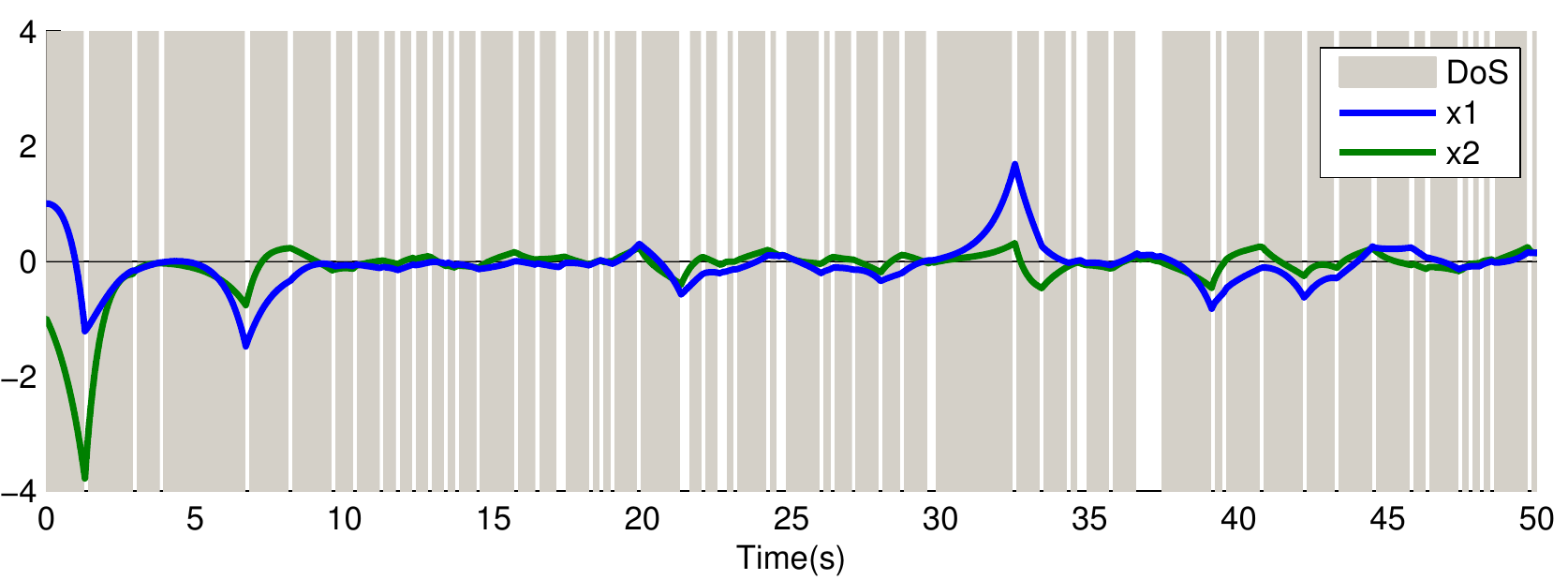} \\
\includegraphics[width=0.45 \textwidth]{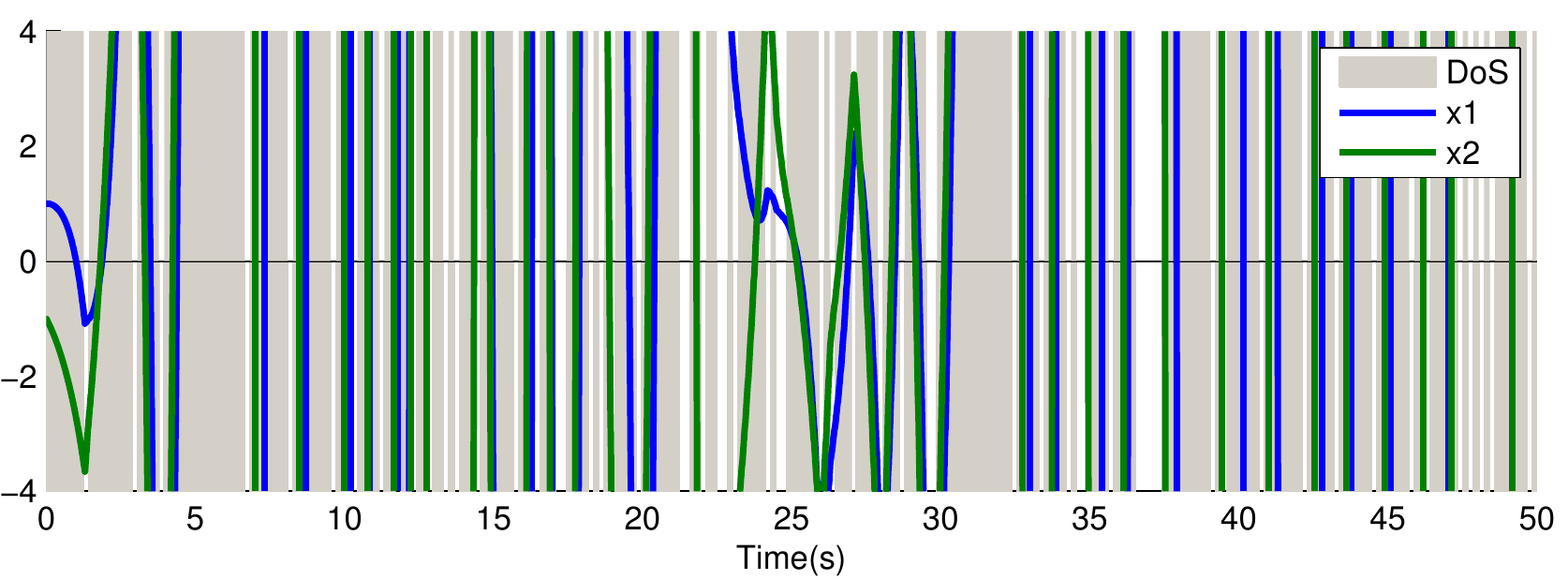} \\
\linespread{1}\caption{Simulation results for the example. 
Top: Analog controller; Center: Digital Controller; 
Bottom: Pure static feedback.} \label{fig:1}
\end{center}
\end{figure}

\section{Example}

The numerical example is taken from \cite{forni}. The system to be controlled
is open-loop unstable and is  
characterized by the matrices 
\begin{eqnarray} \label{eq:system_example}
A=\left[ \begin{array}{cc} \, 1 \, & \, 1 \, \\ 0 \, &  \, 1 \end{array} \right], \quad 
B= \left[ \begin{array}{cc} \, 1 \, & \, 0 \, \\ 0 \, &  \, 1 \end{array} \right]
\end{eqnarray} 
The state-feedback matrix is given by
\begin{eqnarray} \label{eq:controller_example}
K = \left[ \begin{array}{cc} -2.1961 & -0.7545 \\ -0.7545 & -2.7146 \end{array} \right]
\end{eqnarray} 
The control system parameters are $\gamma_1=1$, $\gamma_2=2.1080$, 
$\alpha_1=0.2779$, $\alpha_2=0.4497$, $\|\Phi\|=1.9021$ and $\mu_A=1.5$. 
Disturbance $d$ and noise $n$ 
are random signals with uniform distribution in $[-0.1,0.1]$. 

The network transmission rate is given by $\Delta=0.1$s.
Both analog and digital controllers are considered. 
As for the digital implementation, in accordance with Lemma 3, 
we must select $\sigma$ such that $\sigma < 0.4744$. 
According to (\ref{eq:Delta_bar2}), we obtain the constraint $\delta < 0.1508$.
We select $\delta=0.01$s so that $\delta$ is sufficiently small, and 
in order to synchronize the 
controller sampling rate with $\Delta$.

Figure 1 shows simulation results, which compare
the static feedback law (\ref{eq:static_feedback}) 
with the predictor-based controllers 
(\ref{eq:ct_predictor})-(\ref{eq:state_feedback}) and
(\ref{eq:dt_predictor})-(\ref{eq:state_feedback_dt}). We consider a 
sustained DoS attack with variable period and duty cycle, generated randomly.
Over a simulation horizon of $50$s, the DoS 
signal yields $|\Xi(0,50)|=38.8$s and $n(0,50)=52$. 
This corresponds to values (averaged over $50$s) 
of $\tau_D\approx0.96$ and $T\approx1.29$, and $\sim80\%$ of transmission failures. 
For the predictor-based controllers, the 
stability requirement is satisfied since
\begin{eqnarray}
\frac{\Delta}{\tau_D} + \frac{1}{T} \approx 0.8793
\end{eqnarray}
On the other hand, the DoS parameters do not satisfy the stability 
requirement for the pure static feedback law,
which is  (\emph{cf.} (\ref{tau_lyap1}))
\begin{eqnarray}
\frac{\Delta}{\tau_D} + \frac{1}{T} < 0.0321
\end{eqnarray}
The theoretical bound for the case of pure static feedback is conservative
(indeed, simulations show that (\ref{eq:static_feedback})
ensures closed-loop stability for the system in (\ref{eq:system_example})
up to ${\sim40\%}$ of transmission failures).
Nonetheless, the  improvement given by predictor-based controllers is significant. 
In fact, while the system undergoes instability with (\ref{eq:static_feedback}), 
the  performance level provided by
(\ref{eq:ct_predictor})-(\ref{eq:state_feedback}) and
(\ref{eq:dt_predictor})-(\ref{eq:state_feedback_dt})
is very high despite the sustained DoS attack.

It is worth noting that while stability is independent on the magnitude 
of disturbance and noise signals, performance is not. As shown
in Figure 2, noise significantly impacts on the accuracy of the state estimate, and, hence,
on the closed-loop behavior during DoS status; \emph{cf.} the 
paper conclusions.

\begin{figure}[tb]
\begin{center}
\psfrag{x1}{{\tiny $x_1$}}
\psfrag{x2}{{\tiny $x_2$}}
\psfrag{DoS}{{\tiny DoS}}
\psfrag{ud}{{\scriptsize $u(t_k)$}}
\includegraphics[width=0.45 \textwidth]{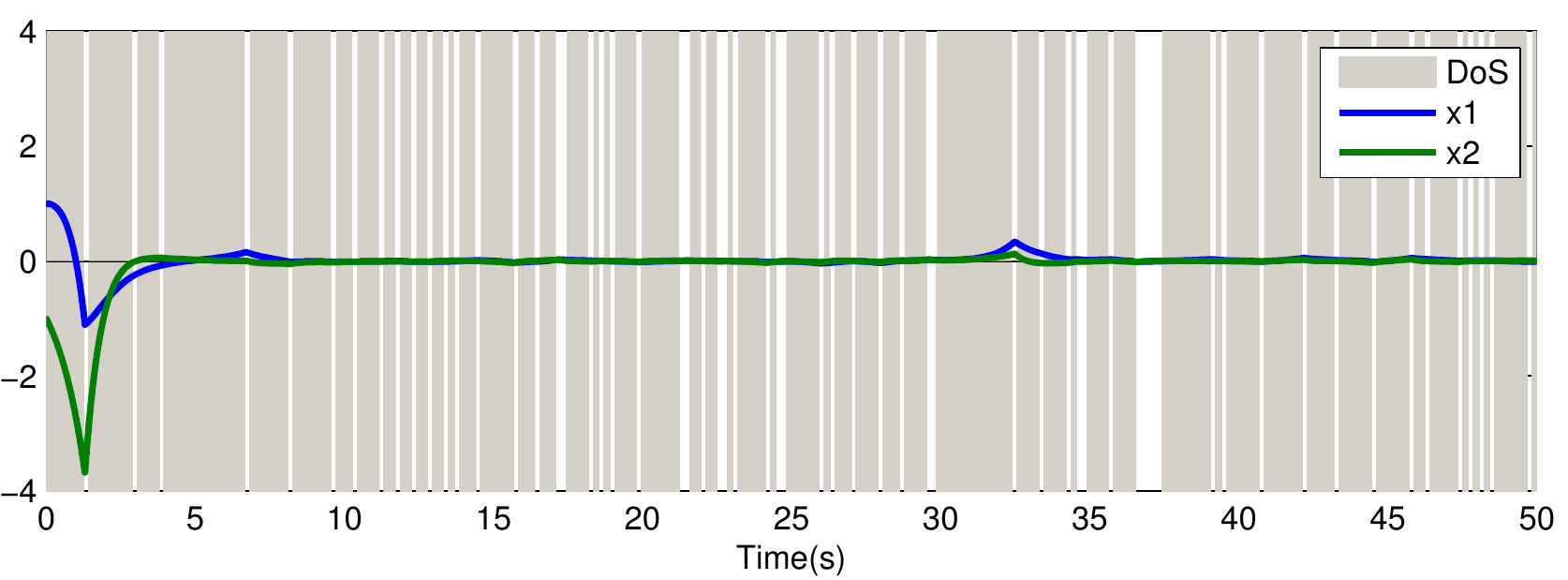} \\
\includegraphics[width=0.45 \textwidth]{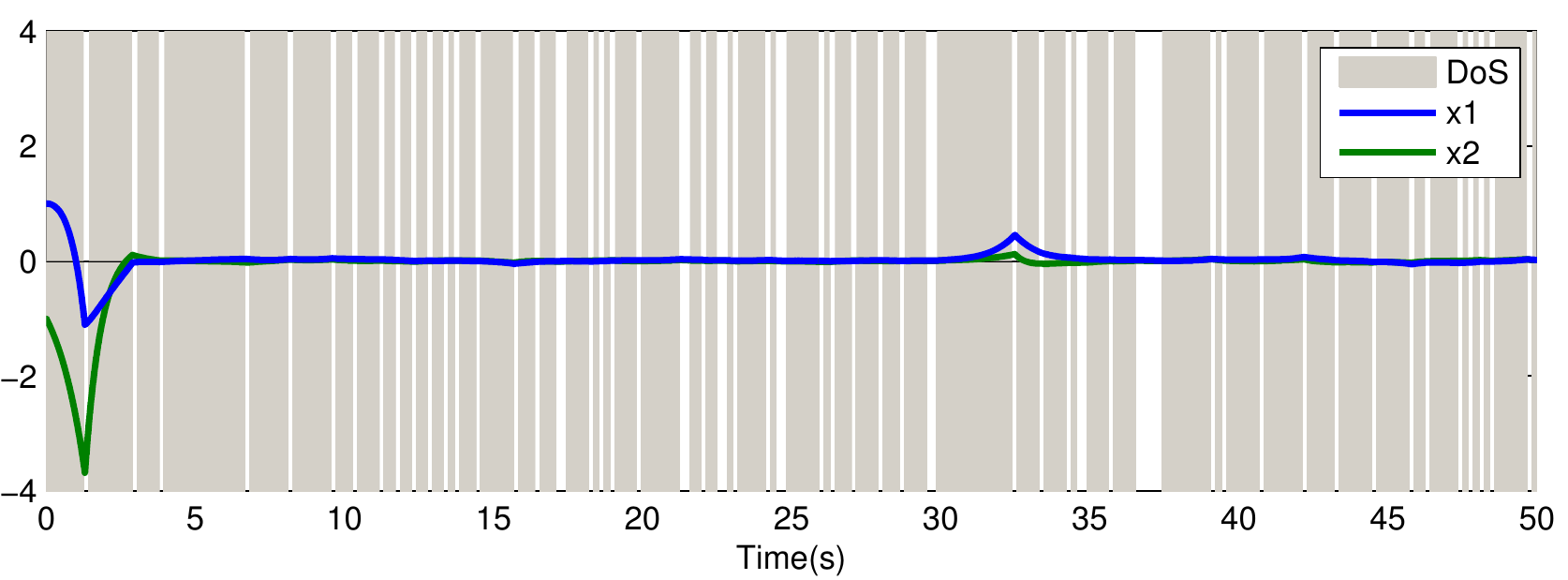} \\
\linespread{1}\caption{Simulation results for the example 
in case disturbance and noise are random signals with uniform distribution in 
$[-0.01, 0.01]$.
Top: Analog controller; Bottom: Digital Controller.} \label{fig:2}
\end{center}
\end{figure}

\section{Concluding remarks}

In this paper, we investigated the problem of designing 
DoS-resilient control systems. It was shown that the 
use of dynamical observers with state resetting 
mechanism makes it possible to 
maximize the amount of DoS that one can tolerate 
for a general class of DoS signals. Both 
analog and digital implementations have been discussed,
the latter requiring a suitable choice of the controller sampling rate.

The results presented in this paper can be extended in various directions.
We envision the use of a similar control architecture for the case 
of partial state measurements, via the approach 
considered in \cite{raff}. Another interesting study
concerns performance robustness against measurement noise,
which is the main factor affecting the quality of the 
process state estimation. The recent results in \cite{Li}
may prove relevant in this regard.

\bibliographystyle{IEEETran}

\bibliography{ref}

\end{document}